\documentclass[aps,prb,nolongbibliography,notitlepage,twocolumn,noeprint,superscriptaddress]{revtex4-2}

\usepackage{amsmath,amssymb,wasysym}
\usepackage{graphicx}
\usepackage{dcolumn}
\usepackage{bm}
\usepackage{braket}
\usepackage{verbatim}
\usepackage{float}
\usepackage{bbold}
\usepackage{color}
\usepackage{xcolor}
\usepackage{relsize}
\usepackage{amsthm}
\usepackage{enumerate}
\usepackage{siunitx}
\usepackage{soul,xcolor}
\usepackage[T1]{fontenc}
\usepackage[colorlinks=true ,urlcolor=blue,urlbordercolor={0 1 1}]{hyperref}

\newcommand{\beq}{\begin{eqnarray}}
\newcommand{\eeq}{\end{eqnarray}}

\newcommand{\bsp}{\begin{split}}
\newcommand{\esp}{\end{split}}

\newcommand{\sgn}{{\rm sgn}}

\newcommand{\be}{\begin{equation}}
\newcommand{\ee}{\end{equation}}

\graphicspath{{./figures/}}
\def\bea{\begin{eqnarray}}
\def\eea{\end{eqnarray}}

\makeatletter

\def\env@sqcases{%
  \let\@ifnextchar\new@ifnextchar
  \left\lbrack
  \def\arraystretch{1.2}%
  \array{@{}l@{\quad}l@{}}%
}
\makeatother

\usepackage{mathtools} 
\begin{document}

\title{Sondheimer magneto-oscillations as a probe of\\ Fermi surface reconstruction in underdoped cuprates}

\author{Alexander Nikolaenko}
\affiliation{Department of Physics, Harvard University, Cambridge MA 02138, USA}

\author{Carsten Putzke}
	\affiliation{Max Planck Institute for the Structure and Dynamics of Matter, 22761 Hamburg, Germany}

\author{Philip J. W. Moll}
	\affiliation{Max Planck Institute for the Structure and Dynamics of Matter, 22761 Hamburg, Germany}
	
\author{Subir Sachdev}
\affiliation{Department of Physics, Harvard University, Cambridge MA 02138, USA}
\affiliation{Center for Computational Quantum Physics, Flatiron Institute, 162 5th Avenue, New York, NY 10010, USA}

\author{Pavel A. Nosov}
\affiliation{Department of Physics, Harvard University, Cambridge MA 02138, USA}

\begin{abstract}
    Determining the Fermi surface (FS) volume in underdoped cuprates is crucial for understanding the nature of the strongly correlated pseudogap phase. Conventional quantum oscillation techniques, typically used for this purpose, are inapplicable in this high-temperature regime due to thermal and disorder-induced smearing of Landau levels. We propose Sondheimer oscillations (SO), semiclassical oscillations of in-plane magnetoresistivity in thin films, as a robust alternative probe of FS reconstruction. SO arise from the commensuration between the cyclotron radius and film thickness, do not rely on Landau quantization, and remain observable at moderate fields and elevated temperatures where quantum oscillations are suppressed. Their frequencies depend solely on the FS parameters (e.g., curvature), and not on specific details of scattering mechanisms. SO are also sensitive to the coherence of inter-layer tunneling, allow contributions from individual FS pockets to be distinguished in the frequency domain, and naturally include the Yamaji angle effect (if present in the system) as a prominent feature in the frequency spectrum. We compute SO spectra as a function of the magnetic field orientation for three representative scenarios: (i) an unreconstructed large FS, (ii) a spin density wave reconstructed FS with volume $p/4$, and (iii) a fractionalized Fermi liquid (FL$^*$) with pocket volume $p/8$ (here $p$ is the hole doping). We show that the SO spectrum offers a wealth of universal features that could be used to differentiate between these scenarios. In particular, we highlight a FS geometry-dependent phase shift between oscillations in longitudinal and transverse conductivities, characterize how the FS curvature can be extracted from SO if the film orientation is perpendicular to the crystallographic $c$-axis, and analyze the evolution of the SO spectrum with doping.
\end{abstract}

\pacs{Valid PACS appear here}
\maketitle
\section{Introduction}

The structure of the Fermi surface (FS) in underdoped cuprate superconductors remains one of the central open questions in the study of high-temperature superconductivity \cite{Timusk1999,Norman2005}. A particularly puzzling aspect is revealed by angle-resolved photo-emission (ARPES) \cite{Norman98,ShenShen05,Johnson11,Shen11,Kondo20,Kondo23,Damascelli25} and scanning tunneling microscopy (STM) \cite{Hoffman14,Davis14} measurements in the under-hole-doped cuprates, which show the emergence of Fermi arcs, with very little, if any, spectral weight on the back-side of an associated hole pocket.

There has been a theoretical debate between different theoretical models of the ARPES/STM data. One class of models has thermally fluctuating $d$-wave superconducting order which yields Fermi arcs over an intermediate temperature range from the broadening of the nodal Bogoliubov quasiparticles \cite{EmeryKivelson,Franz98,Scalapino02,Dagotto05,Berg07,Li_2010,Li11JPhys,Li11PRB,Li_2011,Sumilan17,Majumdar22,YQi23,Xiang24}. A second class of models rely on Fermi pockets with anisotropic spectral weight (discussed below), in which the back-side spectral weight is further suppressed by thermal fluctuations \cite{Pandey25}.

The conventional methods to determine the FS geometry, such as quantum oscillations, fail because they require temperatures well below the cyclotron frequency. However, at such low temperatures, the upper critical field $H_{c2}$ is technically challenging to overcome, making it difficult to probe the normal state. Moreover, in the low temperature and high magnetic field regime, there are clear signatures of broken translational symmetry \cite{Sebastian_review}. A representative  example is the observation of charge order in  YBa$_2$Cu$_3$O$_{6+\delta}$ \cite{LeBoeuf2018}. Therefore, there is still a debate whether the zero field limit has a similar reconstructed Fermi surface. New insight has emerged from recent angle-dependent magnetoresistance (ADMR) measurements \cite{Ramshaw22,Chan2024}, including the evidence for the Yamaji effect \cite{Yamaji1989} in Hg1201 \cite{Chan2024}.
These observations strongly support the presence of hole-like FS pockets with quasiparticles that can tunnel coherently between CuO$_2$ layers, and are consistent with certain pocket FS models of the pseudogap (PG) \cite{Zhao_Yamaji_25,FuChun25}.
It is clearly of interest to have further tests of such FS theories.

In the present work, we argue that observing
Sondheimer oscillations  \cite{Sondheimer1952}  in a cuprate superconductor will provide an even stronger tool to understand the Fermi surface. Sondheimer oscillations (SO) are semiclassical, boundary-induced oscillations of in-plane magnetoresistance that arise in thin films when charge carriers complete an integer number of cyclotron orbits between   consecutive collisions with the sample surfaces. Unlike quantum oscillations that sense extremal orbits with $\partial S/\partial k_z=0$ (here $S$ is the cross-section area enclosed by the quasiparticle trajectory in momentum space), SO provide complementary information about the orbits with $\partial^2 S/\partial k_z^2=0$. Such orbits correspond to the electrons with the highest mobility in the $z$-direction.

Recent experiments have renewed interest in SO across a variety of material systems \cite{Taen2023, Kim2011,Delft2021, Mallik2022, PhysRevB.108.245405,guo2025}. SO are periodic in a magnetic field strength $B$, and do not rely on Landau quantization. As such, they persist at temperatures well above the cyclotron frequency, as long as the mean free path exceeds the film's thickness. This makes SO well-suited for probing the pseudogap regime in cuprate films and for providing important information about coherent interlayer tunneling.

We compute SO frequencies as a function of the magnetic field orientation for three model scenarios: (i) an unreconstructed large FS, (ii) a spin density wave reconstructed FS with volume $p/4$, and (iii) a fractionalized Fermi liquid (FL$^*$) with pocket volume $p/8$ (here $p$ denotes the hole doping).  We show how the Sondheimer frequencies can be used to distinguish between these theoretical proposals and directly extract the FS volume. We further demonstrate that the phase shift between oscillations in the longitudinal and transverse conductivities provides a sensitive probe of FS non-ellipticity. Finally, we analyze the effects of different relative orientations between the crystallographic $c$ axis and the film normal, including a geometry that allows direct extraction of the FS curvature, and we study the doping dependence of the SO spectrum within the FL$^*$ model.

\section{Sondheimer oscillations for arbitrary Fermi surfaces}

In the beginning of this section, we derive Sondheimer oscillations for an arbitrary Fermi surface, following the work of Gurevich \cite{gurevich1959}, and then focus on the elliptical Fermi surface with tight-binding coupling in the $z$-direction, which is  particularly relevant to studying the pseudogap phase of cuprate superconductors.

 \begin{figure*}[t!]
 {\begin{minipage}[h]{0.31\linewidth}
    \center{\includegraphics[width=1\linewidth]{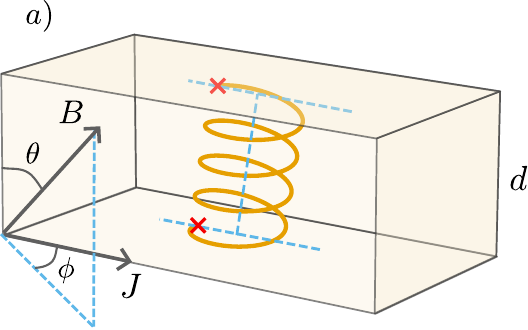}}
    \end{minipage} 
            \hfill
    \begin{minipage}[h]{0.31\linewidth}
    \center{\includegraphics[width=0.9\linewidth]{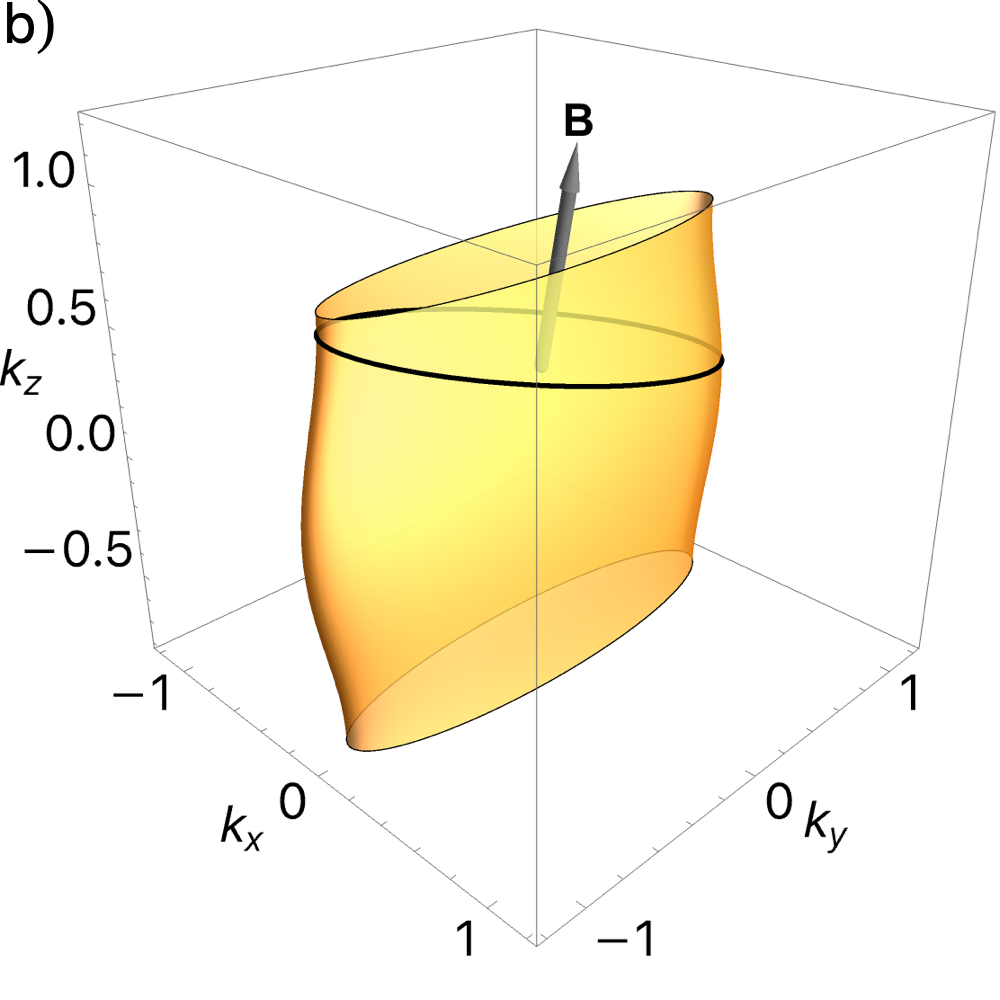}}
    \end{minipage} 
            \hfill
     \begin{minipage}[h]{0.31\linewidth}
    \center{\includegraphics[width=1\linewidth]{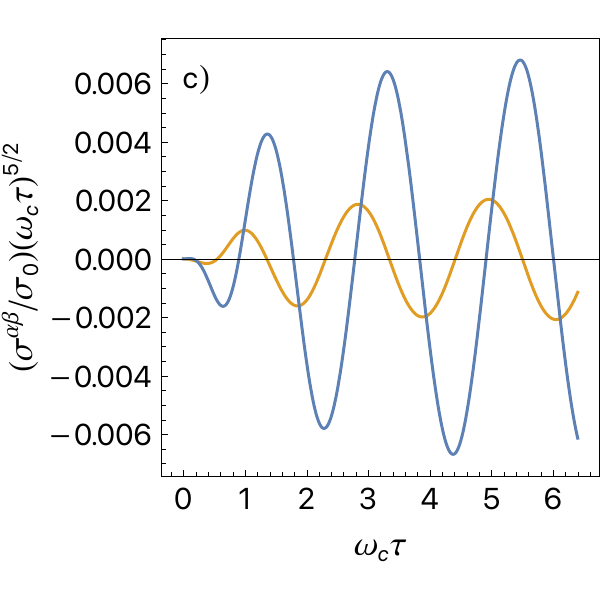}}
    \end{minipage} }
\caption{(a) The trajectory of an electron in the presence of magnetic field. The motion is a superposition of a uniform motion along the direction of the field and spiral rotation around it. Red crosses correspond to surface collisions. (b) The trajectory of an electron in the magnetic field in the Brillouin zone.  The trajectory belongs to a constant energy surface $\epsilon=\epsilon_F$ and lies in the plane, perpendicular to the magnetic field.
	(c) Sondheimer oscillations of conductivity as a function of magnetic field. Blue line corresponds to oscillating component of $\sigma^{xx}$ and orange line to the oscillating component of Hall conductivity $\sigma^{xy}$. They are normalized by zero magnetic field conductivity $\sigma_0^{xx}$ and multiplied by $(\omega_c \tau)^{2.5}$ to ensure that the amplitude saturates to a constant at large magnetic fields. The calculations are performed for an elliptical Fermi surface with magnetic field angles $\theta=20^\circ$, $\phi=45^\circ$, $\tau=0.55ps$ and other parameters given in the caption of Fig.~\ref{fig:frequency_ancilla_elliptic}. }
\label{fig:Fig_1}
\end{figure*}

We consider a thin film of thickness $d$ in the $z$-direction in the presence of the magnetic field $\vec{B}=B( \sin \theta \cos \phi, \sin \theta \sin \phi, \cos \theta)$. The electrons spiral around the magnetic field, while scattering from the surfaces of the film diffusively, see Fig. \ref{fig:Fig_1}(a).  Throughout this work, we treat them semiclassically within the Boltzmann formalism, by introducing the electron distribution function $f(\vec{r},\vec{k})$. When the in-plane electric field $\vec{E}$ is applied, the distribution function is perturbed from its equilibrium value $f=f_0+f'_0(\epsilon)f_1$, where $f_0(\epsilon)=(1+e^{\epsilon/T})^{-1}$ is the equilibrium Fermi-Dirac distribution function. In the linearized regime, the Boltzmann equation in the relaxation-time approximation reads~\cite{Lifshitz1959,gurevich1959}:
\begin{equation}
\omega_c\frac{\partial f_1}{\partial \varphi}+v_z \frac{\partial f_1}{\partial z} +\vec{E}\cdot\vec{v}=-\frac{f_1}{\tau},
 \label{eqn:boltzman_z}
\end{equation}
where $v_\alpha=\partial \epsilon(\vec{k})/\partial k_\alpha$ is the electron velocity and $\epsilon$ is the electron energy.
In this work, we assume that electrons scatter diffusively from the surfaces of the film, which translates into the following boundary conditions $f_1(z=0,v_z>0)=f_1(z=d,v_z<0)=0$. 

We note that in the magnetic field $\vec{B}$ the trajectory of a particle in the Brillouin zone lies on a Fermi surface $\epsilon=\epsilon_F$ and consists of circular motion in the plane perpendicular to the magnetic field, see Fig. \ref{fig:Fig_1}(b). In this case, instead of the Cartesian coordinate system $(k_x,k_y,k_z)$, it is convenient to introduce the coordinates ($\epsilon, \varphi, k_n$), where $\epsilon$ denotes the energy of the particle, $\varphi=\omega_c t$ parametrizes the position along the trajectory and $k_n$ is the momentum component parallel to the magnetic field.  The cyclotron frequency is $\omega_c=B/m^*$ (we work in units where $e=\hbar=1$), and 
the effective mass is $ m^*=(2\pi)^{-1}\partial S(\epsilon,k_n)/{\partial \epsilon}$, where $S(\epsilon,k_n)$ is the area enclosed by the trajectory.

As shown by Gurevich \cite{gurevich1959}, the linearized Boltzmann equation with diffusive boundary conditions can be solved in Fourier space, and the corresponding conductivity can be found. We leave the technical details to Appendix~\ref{app:formalism} and focus here on the final expressions for the conductivity. There are three main contributions that have different origin. The first is the original Chambers formula in the bulk limit \cite{Chambers_1952}. The second is the non-oscillating correction to the bulk conductivity, caused by the finite film thickness \cite{Fuchs1938}. The correction is negative and leads to a reduction in conductivity in thin films. The final one is the oscillatory term leading to Sondheimer oscillations \cite{Sondheimer1952,gurevich1959,nikolaenko2025}. It reads

\begin{equation}
\begin{split}
  \sigma_{\alpha \beta}^{\rm osc}= \int   &  \frac{m^* dk_n d\epsilon d \varphi  }{ (2 \pi)^3\omega_c^2 d}v_\alpha(\varphi)  \frac{\partial f_0}{\partial \epsilon }  \int_{-\infty}^\varphi d\varphi'  |v_z(\varphi')|\\    
 & \int_{-\infty}^{\varphi_1(\varphi')}   d\varphi'' v_\beta(\varphi'')   \exp \left(\gamma(\varphi''-\varphi)  \right),\\        
\end{split}
\label{eq:chambers_mod}
\end{equation}
where $\gamma=1/(\omega_c \tau)$ and $\varphi_1(\varphi)$ determines the phase winding as a particle travels from one surface of the film to the other. Mathematically, it is defined by the following equation
\begin{equation}
    \int_{\varphi_1(\varphi)}^{\varphi}v_z(\varphi')d \varphi'=\pm d \omega_c= \pm u \bar{v}_z,
\end{equation}
where $\bar{v}_z $ is an average $v_z$ along the trajectory and $u=d \omega_c/\bar{v}_z$ is relevant to introduce for future discussion. It corresponds to the total accumulated phase along the trajectory. Note that for certain $\varphi$ the particle may return to the same surface before reaching the opposite one. We omit such trajectories, since they do not contribute to the oscillatory behavior. The velocities $v_\alpha(\varphi)$ could be found numerically by solving two equations     $\omega_c  d \vec{k}/d\varphi= \vec{v} \times \vec{B}$ and $ \vec{v}=\partial_{\vec{k}} \epsilon(\vec{k})$. In the limit of zero temperature, the integration over $\epsilon$ can be performed exactly, and the motion of the electron is restricted to a Fermi surface.

As shown in \cite{nikolaenko2025}, Sondheimer oscillations are largely insensitive to temperature, as long as $T \ll \epsilon_F$ and the relaxation time $\tau$ remains constant, as opposed to quantum oscillations which are exponentially suppressed at $T \gg \omega_c $. The reason is that Sondheimer oscillations are a semiclassical effect: they do not rely on Landau level quantization and therefore persist even when individual Landau levels are thermally broadened and no longer well resolved. This makes Sondheimer oscillations especially relevant in cuprate superconductors, where low temperatures are inaccessible due to the onset of superconductivity or charge order. In practice, however, the mean free path $\bar{l}_z = \bar{v}_z \tau$ is temperature dependent and decreases with increasing temperature. As long as $d < \bar{l}_z(T)$, Sondheimer oscillations remain observable, as discussed below. Importantly, this sensitivity to the temperature dependence of inelastic electron scattering turns Sondheimer oscillations into a unique probe of quasiparticle lifetimes at elevated temperatures. Since the oscillations originate from specific momentum-space trajectories, rather than from a Fermi-surface average as in conventional magnetotransport, they provide access to momentum-resolved information on quasiparticle damping that is otherwise difficult to obtain (see \cite{Delft2021} for a recent application of this method).

Eq.~(\ref{eq:chambers_mod}) can be further simplified in the limit of a strong magnetic field, in which the integration over $k_n$ can be performed using the stationary phase approximation. There are two types of stationary points. The first one consists of end points which correspond to extremal values of $k_n$. For such momenta, the trajectory of an electron in the Brillouin zone reduces to a point. The conductivity oscillates with magnetic field, and the Sondheimer frequency is a function of a Gaussian curvature in the extremal point, tilt of magnetic field $\theta$, and thickness of the film $d$. The oscillations decay as $B^{-4}$ with increasing the magnetic field.

In our setup, this type of oscillations only appears when the magnetic field lies almost parallel to the copper planes, and magnetotransport is strongly suppressed. Therefore, we leave the analysis of this case to the end of the paper, see Section \ref{sec:alternative}.

The second type of stationary points emerges when $u(k_n)=d B/(m^* \bar{v}_z)=\Omega_{\rm SH}(k_n) B$ has an extremum.  Physically, this momentum corresponds to electrons winding down the extremal phase while traveling between two surfaces.  Note that at such momenta, the area enclosed by the trajectory in the Brillouin zone is not extremal as it is in quantum oscillations. Rather, the rate of change of the area is extremal. This follows from the relation $m^* \bar{v}_z=(1/2\pi)\partial S/\partial k_z$, as shown in \cite{Harrison1960}. This property makes Sondheimer oscillations a complementary measurement to quantum oscillations, since it provides information about non-extremal portions of the Fermi surface. Generally, oscillations of this type are non-sinusoidal, with the first harmonic given by:


 \begin{equation}
 \begin{aligned}
 \sigma_{\alpha \beta}^{\rm osc}&=  \frac{m^* \tau^2 |c^{(1)}_{\alpha \beta}(k^*)|}{4\pi^3d (1+  \omega_c^2 \tau^2)}   e^{-d/(|\bar{v}_z| \tau)} \times  \\
 &\times \sqrt{\frac{2 \pi  }{  |u''(k^*)|}} \cos\left(   \Omega_{\rm SH} B +\lambda_{\alpha \beta}\right)\Bigg|_{k_n=k^*},   \label{eq:oscil_full_sp}
 \end{aligned}
\end{equation}
where $c^{(s)}_{\alpha \beta}(k^*)$ are Fourier components $ c^{(s)}_{\alpha \beta}(k_n)=\int ds/(2 \pi) e^{i su}c_{ \alpha \beta}(k_n,u)$ and 
\begin{equation}
c_{\alpha \beta }(k_n,u) =  \int \bar{v}_z^2   d \varphi    \frac{  v_\alpha(\varphi) v_\beta(\varphi_1(\varphi))  }{v_z(\varphi_1(\varphi))} 
\end{equation}
is a periodic function of $u$ with a period $2 \pi$. The phase
\begin{equation}
\lambda_{\alpha \beta}=\arg(c^{(1)}_{\alpha \beta}(k^*)) + \frac{\pi}{4} \sgn(u''(k^*))+ \tan^{-1} \left( \frac{2 \omega_c \tau}{1-\omega_c^2 \tau^2}\right).
\end{equation}
Let us analyze Eq.~(\ref{eq:oscil_full_sp}) in more detail. First, the oscillations are exponentially suppressed by a factor $e^{-d/\bar{l}_z}$, where $\bar{l}_z=\bar{v}_z \tau$. In thick films with $d \gg \bar{l}_z$ electrons no longer traverse ballistically between the boundaries; impurity scattering rapidly randomizes their velocities, and the memory of the boundaries is lost in the bulk. As a result, the oscillations quickly disappear. Second, the frequency of oscillations is given by Sondheimer frequency $\Omega_{\rm SH}=d/(m^*\bar{v}_z)$ evaluated at the extremum momentum $k_n=k^*$.
To observe multiple oscillation periods, the following inequality should be satisfied $(\omega_c \tau) d/\bar{l}_z \gg 1$. It means that oscillations should be clearly visible in the $\omega_c \tau \gg1$ regime of strong magnetic fields.

Finally, contrary to the first type of oscillations, the second type decays as a power law $B^{-5/2}$ in the magnetic field.

\subsection{Application to quasi-2D elliptical Fermi surfaces}

After reviewing the general formalism, we turn our attention to the cuprate materials. 
Most of them have copper oxide planes that are weakly hybridized. In this section, we consider the copper oxide planes coinciding with the $xy$ planes of our geometry, see Fig.~\ref{fig:alternative_geometry}(a), while exploring alternative geometries in the next section.

\begin{figure*}[t!]
 
 \begin{minipage}[h]{0.3\linewidth}
\center{\includegraphics[width=1\linewidth]{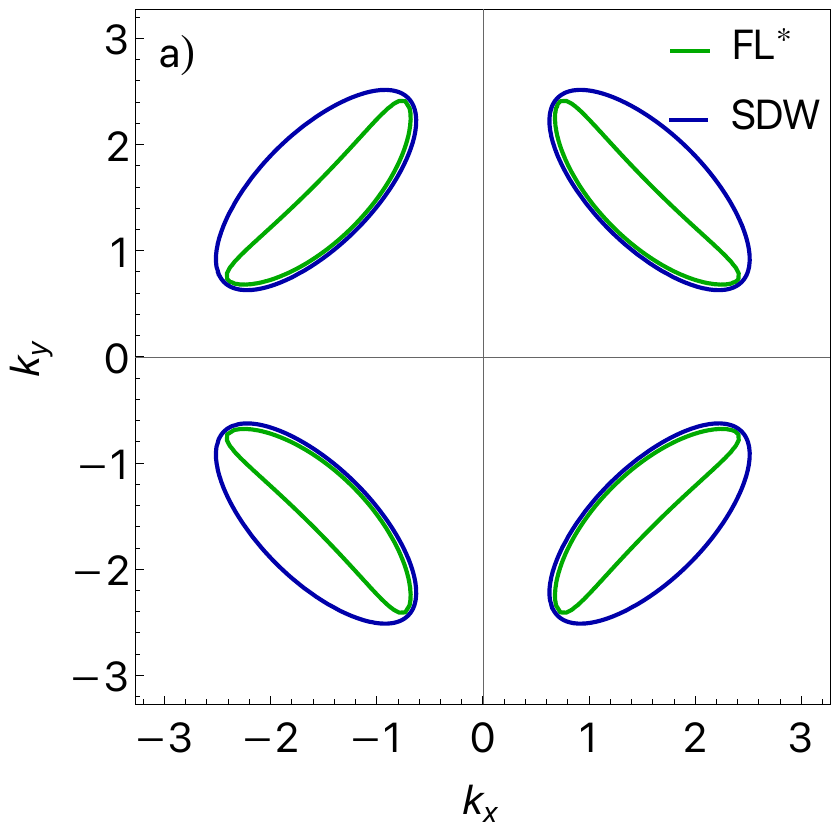}}
 \end{minipage}
       \hfill
    \begin{minipage}[h]{0.3\linewidth}
    \center{\includegraphics[width=1\linewidth]{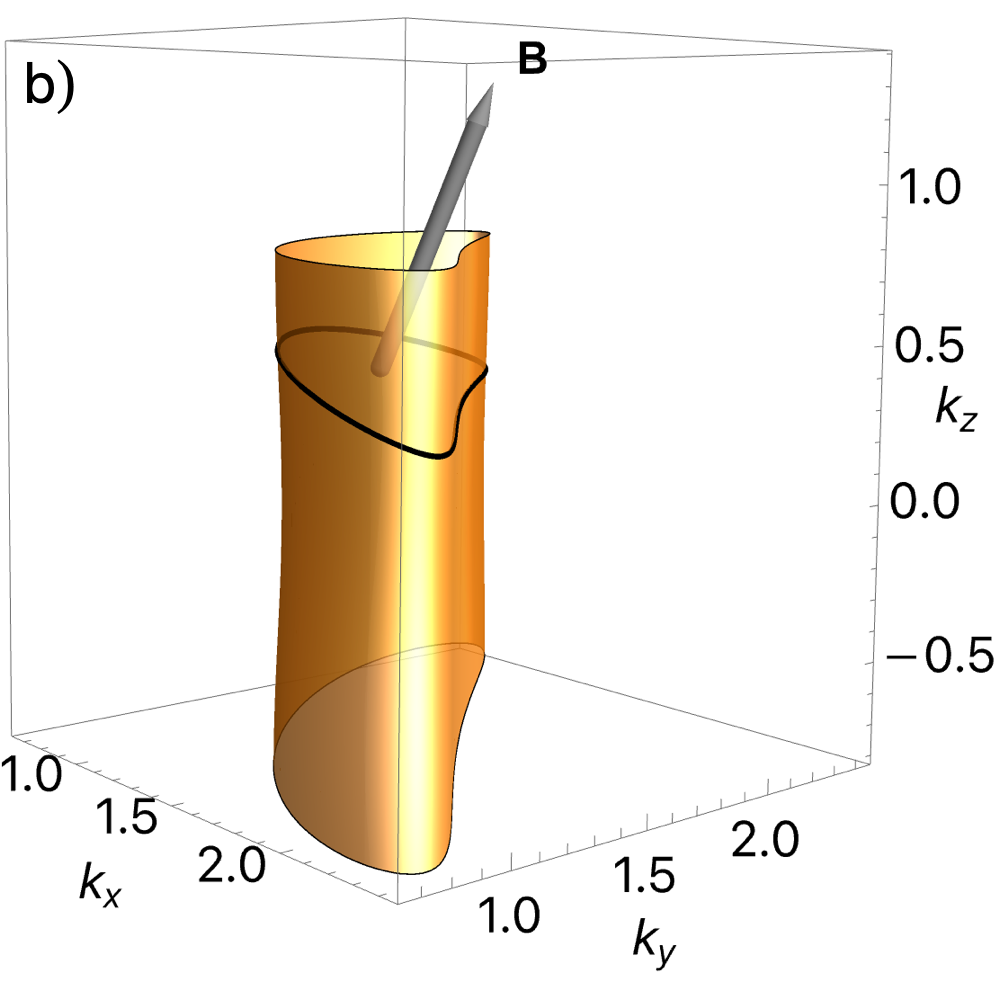}}
    \end{minipage} 
            \hfill
     \begin{minipage}[h]{0.3\linewidth}
    \center{\includegraphics[width=1\linewidth]{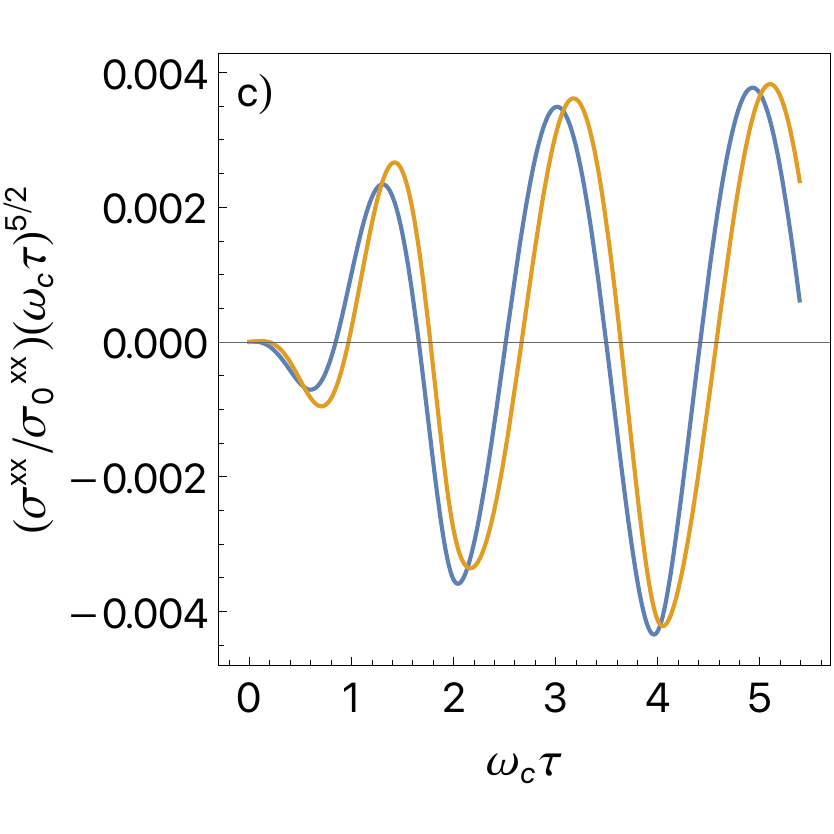}}
    \end{minipage} 
\caption{(a) The green line shows the dispersion of the FL$^*$ model with small hole pockets of area $p/8$ while the blue line shows the dispersion of the SDW hole pockets with the area $p/4$. (b) The trajectory of an electron in the magnetic field in the Brillouin zone.  The trajectory belongs to a constant energy surface $\epsilon=\epsilon_F$ of the FL$^*$ dispersion, derived from Eq.~(\ref{eq:FLs}).
	(c) Sondheimer oscillations of conductivity as a function of magnetic field. Blue line corresponds to the oscillating component of $\sigma^{xx}$ and orange line to the oscillating component of Hall conductivity $\sigma^{xy}$. They are normalized by zero magnetic field conductivity $\sigma_0^{xx}$ and multiplied by $(\omega_c \tau)^{5/2}$ to ensure that the amplitude saturates to a constant at large magnetic fields. The calculations are performed for magnetic field angles $\theta=20^\circ$, $\phi=45^\circ$, the hopping in the $z$-direction $t_\perp=6m eV$, $\eta=0$, the thickness of the film $d/a=40$, the relaxation time $\tau=0.55 ps$, and using the FL$^*$ Fermi surface parameters given in the Appendix \ref{app:ancilla}.}
\label{fig:Fig_ancilla}
\end{figure*}

In the most challenging and theoretically interesting underdoped phase, the Fermi surfaces of cuprate superconductors resemble four small hole pockets centered on the diagonal in the 2D $(k_x,k_y)$ plane, see Fig.~\ref{fig:Fig_ancilla}(a).  In this subsection, we approximate them with elliptical Fermi surfaces, while discussing a more general case in the next section. Since the planes are weakly hybridized in the $z$-direction, the tight-binding model is sufficient to describe the $k_z$ dispersion.

Ultimately, we consider electrons with the following dispersion:
\begin{equation}
    \epsilon=\frac{k_x^2}{2m_x}+\frac{k_y^2}{2m_y}-2 t_\perp \cos (c k_z),
    \label{eq:disp_elliptic}
\end{equation}
where $c$ is the lattice constant and $t_\perp$ is a tight-binding hopping parameter in the $z$-direction, and $m_x$ and $m_y$ determine the size of the hole pocket. We further assume that $t_\perp \ll (m_{x,y} a^2)^{-1}$ where $a$ is the lattice constant in the $xy$ plane. This is true for most of the cuprates, since the planes are only weakly hybridized. 

\begin{figure}[t!]
       \centering
		\includegraphics[width=0.8\linewidth]{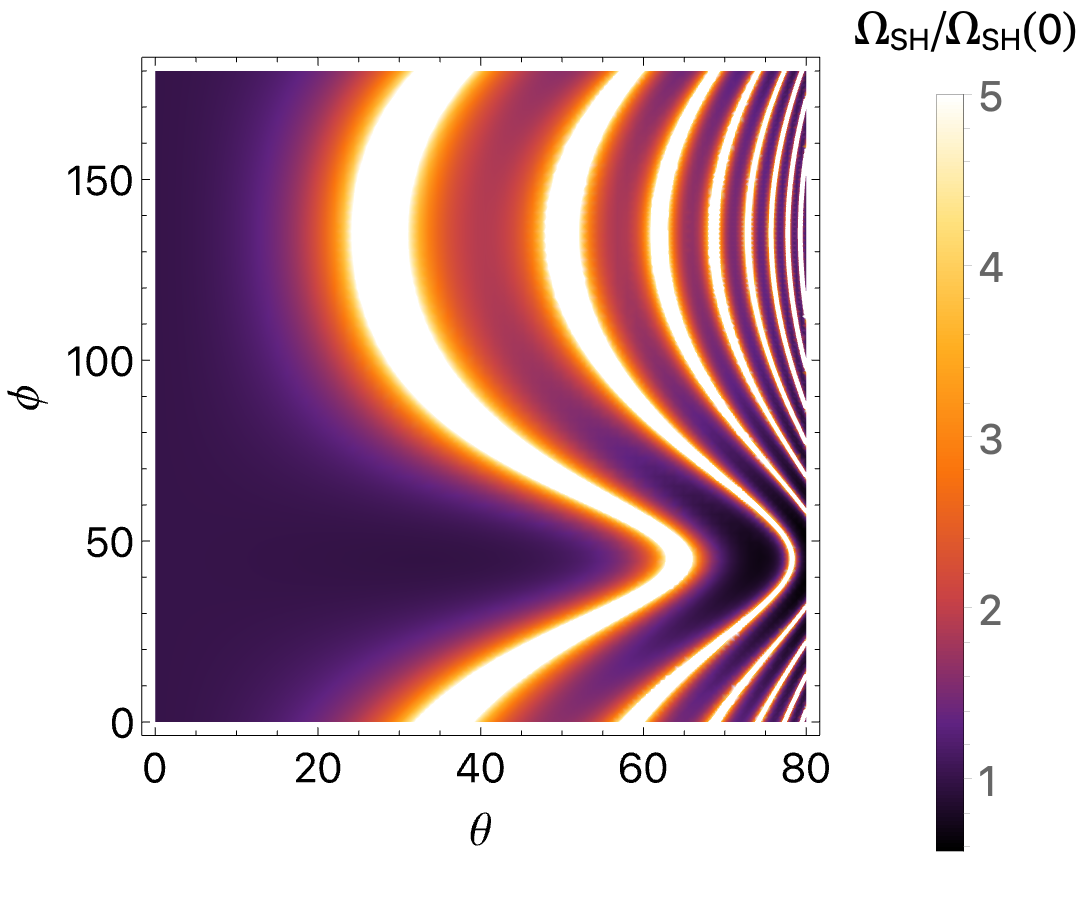}
		\caption{Sondheimer frequency for an elliptic Fermi surface with the dispersion in Eq.~(\ref{eq:disp_elliptic}). Parameters: $m_x a^2/\hbar^2=0.82 eV^{-1}$, $m_y a^2/\hbar^2=13.3eV^{-1}$, $\epsilon_F=0.05 eV$, $t_\perp=3meV$ and $c/a=4$ and $d/a=40 $, with $a= \SI{3.9 }{\angstrom}$. Such parameters correspond to approximately 10 layers in the $z$-direction and the effective mass $m^*\approx1.66m_e$ at $\theta=0$. The plot is normalised by $\Omega_{\rm SH}(0)=0.12 \, T^{-1}$, which is the Sondheimer frequency at $\theta=0$ when the magnetic field is perpendicular to the thin film.}
	\label{fig:frequency_ancilla_elliptic}
\end{figure}

In this special case most of the progress can be made analytically. The effective mass $m^*=\sqrt{m_x m_y}/\cos \theta$ does not depend on $k_n$ and is only a function of $\theta$. The extremal trajectory thus corresponds to the maximum $\bar{v}_z$ which happens to be at $\bar{k}_z c = \pi/2$. Such a trajectory is shown in Fig.~\ref{fig:Fig_1}(b). The Sondheimer frequency is 

\begin{equation}
    \Omega_{\rm SH}(\theta,\phi)=\frac{d  \cos \theta}{2 t_\perp c \sqrt{m_x m_y} J_0\left( k_{\rm cal}(\phi) c\tan \theta \right)}\;,
    \label{eq:sondh_ell}
\end{equation}
where $J_0(x)$ is the  Bessel function of the first kind and 
\begin{equation}
    k_{\rm cal}(\phi)=\sqrt{2 \epsilon_F(m_x \cos^2 \phi+m_y \sin ^2 \phi) } 
\end{equation}
is the caliper radius of an ellipse. We emphasize that the Sondheimer frequency does not depend on the relaxation time $\tau$, and thus provides a clean probe of a material's band structure. By rotating the magnetic field and detecting a change in Sondheimer frequency, one can find all the necessary parameters and fully characterize the Fermi surface. For example, computing the second derivative with respect to $\theta$ gives direct access to the caliper radius, without involving any other parameters, such as $t_\perp$:

\begin{equation}
\left.\frac{\partial_{\theta}^{2}\Omega_{\rm SH}}{\Omega_{\rm SH}}\right|_{(0,\phi)}
= \frac{k_{\rm cal}(\phi)^{2} c^{2} - 2}{2}.
\label{eq:caliper}
\end{equation}

Note that when the angle of the magnetic field is
\begin{equation}
    \theta_{\rm Yamaji}=\arctan\left(\frac{\pi(n-1/4)}{k_{\rm cal}(\phi)c}\right),
    \label{eq:yamaji}
\end{equation}
where $n\in \mathrm{Z}$, then the $\bar{v}_z$ component of the velocity goes to zero, and the enclosed areas $S(\epsilon, k_n)$ do not depend on $k_z$ (these two facts are tied together by the relation $m^*\bar{v}_z \sim \partial S/\partial k_z$ \cite{Harrison1960,kartsovnik2004}). This effect was first discovered by Yamaji \cite{Yamaji1989}, who showed that the out-of-plane resistivity should be substantially increased at such angles due to the lack of coherent $z-$axis transport. Recent measurements of the out-of-plane resistivity in the pseudogap phase of the single-layer cuprate superconductor Hg1201 have provided evidence consistent with the Yamaji effect \cite{Chan2024}. The resistivity displayed pronounced peaks at the Yamaji angles of the magnetic field, which made it possible to estimate the size of the small hole pockets inside the pseudogap phase.

In this work, we offer another consequence of the Yamaji effect, namely the disappearance of the oscillations, accompanied by a rapid increase of their frequency at $\theta\sim \theta_{\rm Yamaji}$. In particular, from Eq.~\eqref{eq:oscil_full_sp} we find that the oscillation amplitude decays exponentially near the Yamaji angle as
\begin{equation}
   \sigma^{\rm osc}_{\alpha\beta}\propto \exp\left({-d/(2 t_\perp c^2 k_{cal}(\phi) \tau |\delta \theta|}) \right)\;,
\end{equation}
where $\delta \theta=\theta-\theta_{\rm Yamaji}$ and $\phi$ is fixed. By measuring this suppression, one can extract quantitative information regarding the interlayer tunneling amplitude and the caliper radius. This behavior is also accompanied by the divergence of the Sondheimer frequency, see Fig.~\ref{fig:frequency_ancilla_elliptic}. 
The detection of these features offers a direct means of determining the size of the hole pockets.



 A typical oscillatory contribution to the conductivity is shown in Fig.~\ref{fig:Fig_1}(c). The amplitude of such oscillations is of the order of a tenth of a percent compared to the bulk value, which makes them within the resolution of a physical device. We can also distinguish between the Hall and longitudinal conductivities by the phase shift of the oscillations: in the case of an elliptical Fermi surface, it is equal to $\pi/2$. That is because the $v_x$ and $v_y$ components of the velocity are shifted by a phase of $\pi/2$  relative to each other. In the next section, we will see that for more general, non-elliptic Fermi surfaces, this condition no longer holds.

\section{FL$^*$ and SDW model}
Having established the general behavior using a simplified elliptic model, we now turn to the more concrete theoretical descriptions of cuprate superconductors. 
\subsection{Underdoped case}

We start with the underdoped phase, commonly referred to as the pseudogap phase. While the precise physics behind this state is still under debate, several competing theories were proposed to describe the fascinating features of this state, such as the emergence of the Fermi arcs. Here we focus on two such frameworks, namely the spin-density-wave (SDW) \cite{SchmalianPines1,SchmalianPines2,Tremblay04,Shen11,Chubukov23,Chubukov25} and fractionalized Fermi liquid (FL$^*$)
\cite{TSSSMV03,Si2004,TSSSMV04,APAV04,GS10,Qi10,SSMetlitskiPunk12,Bonderson16,Vojta20,Coleman22,Tsvelik24,Tsvelik25} models.

The first one relates the Fermi arcs to the emergence of the magnetic order and the subsequent reconstruction of the Fermi surface. In the SDW phase, the Hamiltonian is given by

\begin{equation}
    H_{\rm SDW}=\epsilon_c(\mathbf{k})c^{\dag}(\mathbf{k})c(\mathbf{k})+\Delta  c^{\dag}(\mathbf{k})c(\mathbf{k+Q})+h.c.,
    \label{eq:Ham_SDW}
\end{equation}
where $\epsilon_c(\mathbf{k})$ is the electronic dispersion in the overdoped case, $Q$ is the wave-vector of the SDW phase, which we assume to be $Q=(\pi,\pi)$ throughout this work (antiferromagnetic ordering), and $\Delta $ is an order parameter. As a result of the hybridization, the large Fermi surface reconstructs, and small hole pockets appear, see Fig.~\ref{fig:Fig_ancilla}(a). The area of each pocket is $p/4$, where $p$ is the doping. As in the previous section, the coupling between the layers is assumed to be $-2 t_\perp \cos( c k_z)$. The Hamiltonian in Eq. (\ref{eq:Ham_SDW}) can be diagonalized and the Sondheimer frequency can be found at different orientations of the magnetic field. 
Using the Sondheimer frequency, one can reconstruct the full Fermi surface by rotating the magnetic field. The first measurement can be performed at small $\theta$. While Eq.~(\ref{eq:caliper}) strictly does not hold for general Fermi surfaces, one can use its r.h.s. to {\it define} a proxy $k_{\rm proxy}(\phi)$ for the actual caliper radius, i.e. $c^2k^2_{\rm proxy}(\phi):= 2(1+\Omega_{\rm SH}^{-1}\partial_\theta^2\Omega_{\rm SH})|_{\theta=0}$. For Fermi surfaces that are close to an elliptical shape $k_{\rm proxy}(\phi)\approx k_{cal}(\phi)$. For the SDW Fermi surface, given by the blue line in Fig.~\ref{fig:Fig_ancilla}(a), the two momenta almost coincide, allowing for Fermi surface detection using nearly perpendicular magnetic fields. 



 \begin{figure}[t!]
    \begin{minipage}[h]{0.6\linewidth}
    \center{\includegraphics[width=1\linewidth]{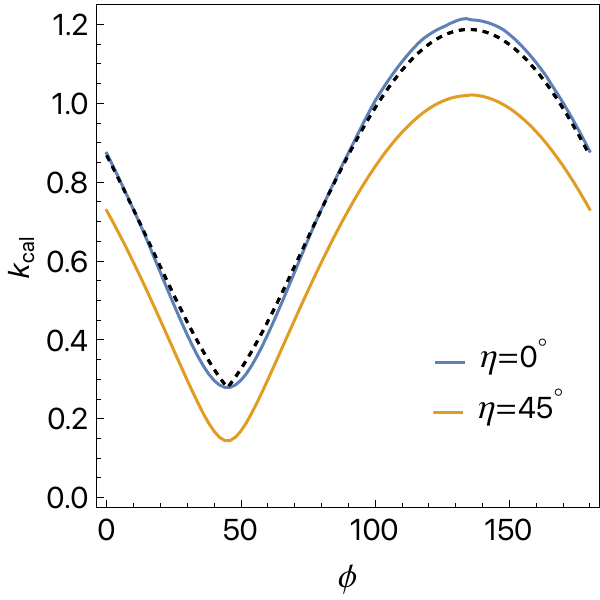}}
    \end{minipage}            
\caption{The black dashed line shows the caliper radius of FL$^*$ Fermi surface (the green line in Fig.~\ref{fig:Fig_ancilla}(a)) as a function of the angle $\phi$. The orange and blue lines show the $k_{\rm proxy}(\phi)$ radius extracted from the second derivative of the Sondheimer frequency for different values of the parameter $\eta$ controlling the hopping in the $z$-direction. Parameters: $t_\perp=3meV$, $d/a=40$. }
\label{fig:Fig_caliper}
\end{figure}
\begin{figure*}[t!]
 \center{\includegraphics[width=1\linewidth]{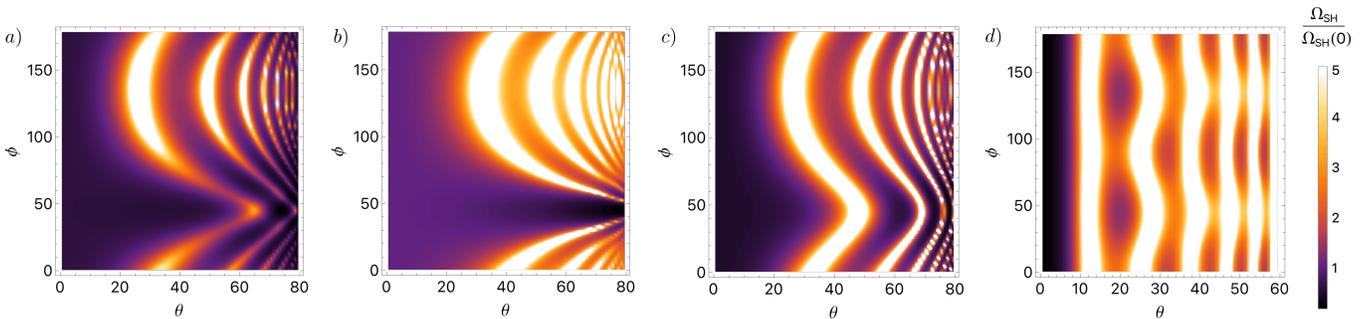}}
    
\caption{Plots of the Sondheimer frequency in the underdoped case for the FL$^*$ model with: (a) $\eta=0^\circ$,  (b) $\eta=45^\circ$, and SDW model with $\Delta=0.14 eV$ (c). $\Omega_{\rm SH}(0)$ is Sondheimer frequency at $\theta=0$ when the magnetic field is perpendicular to the thin film. Parameters: $t_\perp=3 meV$,$d/a=40$. Panel (d) shows the Sondheimer frequency in the overdoped case $\Phi=0$. }
\label{fig:frequency_ancilla}
\end{figure*}

Another measurement can be done by rotating the magnetic field away from $\theta=0$. Fig. \ref{fig:frequency_ancilla}(c) shows the Sondheimer frequency as a function of two angles $(\theta,\phi)$. The divergences occur at the Yamaji angles, which closely follow the predictions of Eq. (\ref{eq:yamaji}). The detection of the Yamaji angles provides another way to find the caliper radius $k_{\rm cal}(\phi)$ at every orientation $\phi$ and allows one to fully reconstruct the Fermi surface.

While the SDW hypothesis naturally explains the reconstruction of the Fermi surface, it requires the presence of magnetic order (finite $\Delta$). Such order is detected by neutron scattering experiments only at very low dopings.

Another theory that captures the key features of the pseudogap phase is an FL* state. In this scenario, strongly renormalized electronic quasiparticles coexist with the spin-liquid background. One concrete realization of the FL* model, namely the Ancilla layer model, has been shown to successfully reproduce many features of the pseudogap phase, including Fermi arcs and quantum critical behavior~\cite{Zhang2020,Zhang2020_2,Mascot2022,Nikolaenko2023,Christos2024,BCS24,Boulder25}.

We omit the possible contribution from the spin-liquid sector in this work and focus exclusively on the renormalized electronic degrees of freedom. In the simplest mean-field description, the Hamiltonian takes the form
\begin{equation}
\begin{split}
    H_{FL^*}=\epsilon_c&(\mathbf{k})c^{\dag}(\mathbf{k})c(\mathbf{k})+\epsilon_f(\mathbf{k})f^{\dag}(\mathbf{k})f(\mathbf{k})+\\
   +&\Phi  c^{\dag}(\mathbf{k})f(\mathbf{k})+\Phi^*  f^{\dag}(\mathbf{k})c(\mathbf{k}),\\
    \end{split}
    \label{eq:FLs}
\end{equation}
where $\epsilon_c(\mathbf{k})$ is the dispersion of physical electrons, $\epsilon_f(\mathbf{k})$ is the dispersion of the hidden electrons, and $\Phi$ is the hybridization between the two Ancilla layers. 
As a result of the hybridization, the electron Fermi surfaces acquire a gap in the antinodal region, and the hole-pockets emerge, see Fig.~\ref{fig:Fig_ancilla}(a). The parameters are chosen as in Ref.~\cite{Mascot2022}, see Appendix \ref{app:ancilla} for details.

Crucially, because the second layer consists of spinons and a background spin-liquid, the Luttinger relation is modified. As a result, the size of the hole pockets is reduced from $p/4$ to $p/8$. This distinction in the size of the Fermi surface is an important checkpoint in establishing a correct theory of the pseudogap state.

The hopping in the $z$-direction can be described in the following way. Let us introduce a physical electron $c_0(\mathbf{k})=(\cos \eta) c(\mathbf{k})+(\sin \eta) f (\mathbf{k})$. The hopping $H_z=-2 t_\perp \cos (k_z c) c^\dag_0(\mathbf{k})c_0(\mathbf{k})$ should be added to the  $H_{FL^*}$ Hamiltonian and be incorporated in the full 3D theory. A typical trajectory of a quasiparticle in the presence of the magnetic field is shown in Fig.~\ref{fig:Fig_ancilla}(b).

We first compute the Sondheimer frequency at small angles $\theta$ and estimate the caliper radius by computing the second derivative, see Eq.~(\ref{eq:caliper}).  Fig.~\ref{fig:Fig_caliper} shows that this method works well even for more general dispersion relations, and one can clearly understand the orientation of the small pockets in the Brillouin zone and obtain a good estimate of their size.

In Fig. \ref{fig:frequency_ancilla}(a),(b) we compute the Sondheimer frequency for the Ancilla model, with Hamiltonian $H_{FL^*}+H_z$  at different magnetic field orientations and compare it to the SDW scenario. We see that the Yamaji angles are shifted, with the shift appearing most noticeable at angles close to $\phi=\pi/4$. This is expected, as the caliper radius in this direction is approximately half that of the SDW scenario. Furthermore, there is a notable disappearance of Yamaji peaks at $\eta=45^\circ$ close to $\phi=45^\circ$. At these angles, the dispersion deviates from the elliptical approximation significantly, and the Yamaji predictions no longer hold. A similar disappearance of the Yamaji peaks was reported in \cite{zhao2025}.

The typical oscillations of conductivity for the FL$^*$ model are shown in Fig.~\ref{fig:Fig_ancilla}(c). Additionally, it is possible to extract more information from the relative phase of Hall and longitudinal response. In the case of an elliptical Fermi surface, it is equal to $\pi/2$, while for the FL$^*$ model it is close to zero, see Fig.~\ref{fig:Fig_1}(c). This is because $v_x$ and $v_y$ components of the velocity are no longer shifted by $\pi/2$ but rather almost coincide with each other.

\subsection{Overdoped case}

So far, we have been focusing on the underdoped regime, where the presence of the pseudogap phase is most readily established. However, the nature of the transition from the underdoped to the overdoped phase presents a mystery as well. In particular,  experiments on a wide range of materials report signatures of a carrier-density transition from $p $ to $1+p$, as evidenced by a change in the Hall number \cite{Badoux2016, Putzke2021}.

We start by discussing the overdoped case, where Sondheimer oscillations may be more readily observed. The overdoped side is well described by the Ancilla formalism: the second and third layers form spin singlets and effectively decouple from the system, while the first layer forms a large Fermi surface described by the dispersion $\epsilon_c(\mathbf{k})$. 

Fig.~\ref{fig:frequency_ancilla}(d) shows the Sondheimer frequency for a large Fermi surface as a function of two angles $(\theta, \phi)$. In comparison to Fig.~\ref{fig:frequency_ancilla}(a),(b), the Yamaji peaks occur at smaller angles $\theta$. This is clear from Eq.~(\ref{eq:yamaji}): the caliper radius of the large Fermi surface increases and shifts the Yamaji peaks toward smaller angles $\theta$.

\begin{figure}[t!]
    \center{\includegraphics[width=0.7\linewidth]{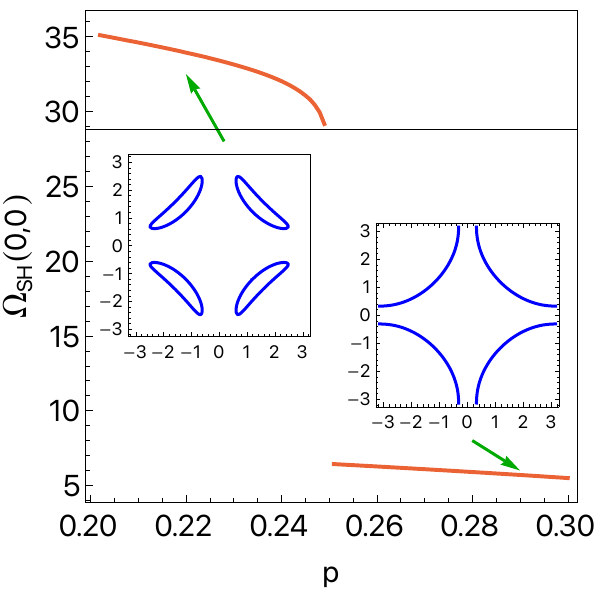}}
\caption{Sondheimer frequency at a perpendicular magnetic field ($\theta=0$) as a function of doping across the transition from the overdoped to the underdoped side. Parameters: $t_\perp=3\times 10^{-3} eV$, $\eta=0$, $d/a=40$.}
\label{fig:Fig_overdoped}
\end{figure}

Next, we discuss the evolution of the Sondheimer frequency across the transition from the underdoped to the overdoped side. In the simplest scenario, the transition could be modeled as a mean-field solution with doping-dependent hybridization $\Phi(p)=\Phi_0 \sqrt{p-p_c}$. We choose $p_c=0.25$, which is close to experimental findings, and determined $\Phi_0$ from the value on the underdoped side. Fig.~\ref{fig:Fig_overdoped} shows the change in the Sondheimer frequency across the transition. As the Fermi surface volume reconstructs, the Sondheimer frequency experiences a finite jump (which could be further smeared by the thermal and quantum fluctuations beyond the mean-field description). The observation of such a rapid reconstruction could provide useful information about the nature of this transition.

\section{Alternative film geometry}
\label{sec:alternative}

 \begin{figure*}[t!]
    \begin{minipage}[t!]{0.3\linewidth}
    \center{\includegraphics[width=0.65\linewidth]{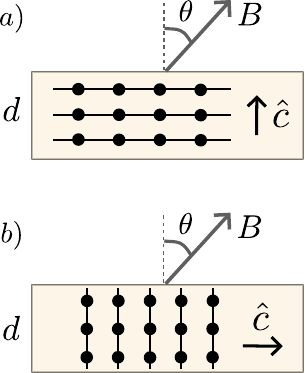}}
    \end{minipage} 
\hfill
    \begin{minipage}[h]{0.35\linewidth}
    \center{\includegraphics[width=1\linewidth]{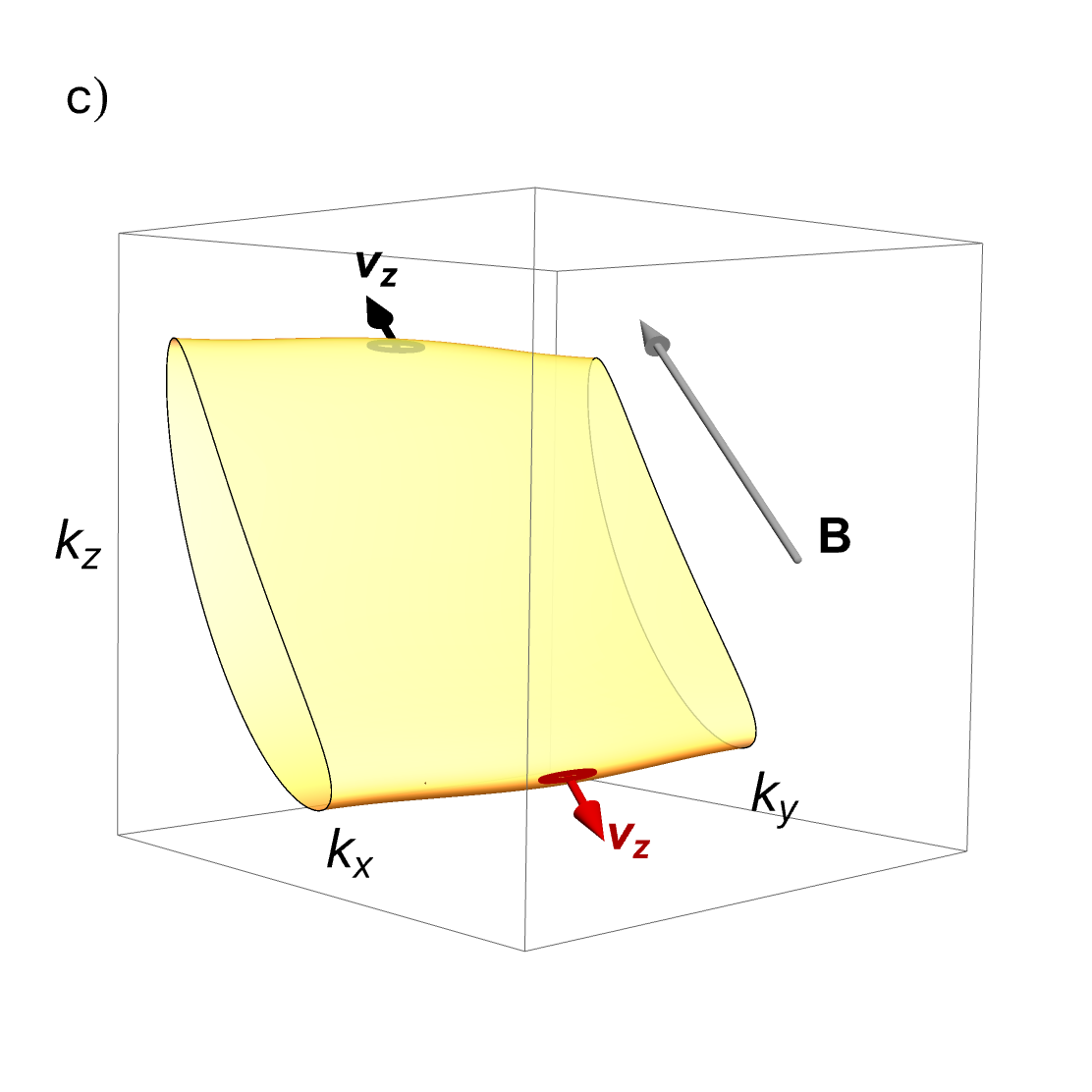}}
    \end{minipage}     
    \hfill
    \begin{minipage}[h]{0.28\linewidth}
    \center{\includegraphics[width=1\linewidth]{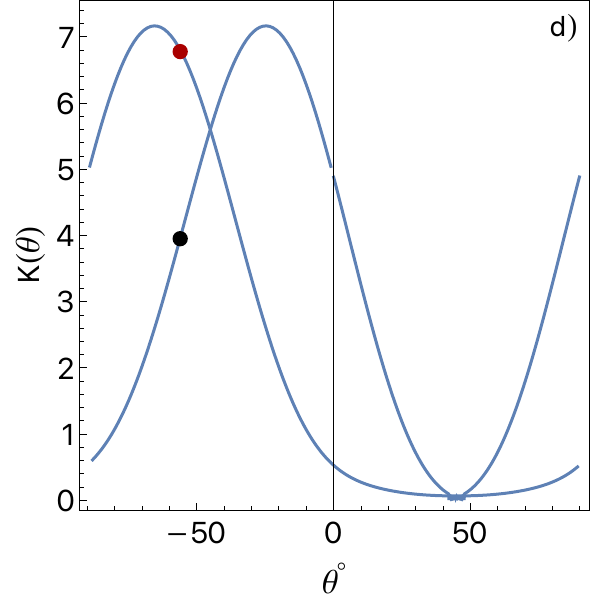}} 
    \end{minipage}  
\caption{a) A setup where the crystallographic $\hat{c}$-axis is perpendicular to the film boundary. Black lines denote copper planes. b) Alternative geometry where the crystallographic $\hat{c}$-axis is parallel to the film boundary. c) The boundary points in the Brillouin zone for the alternative geometry setup. d) Gaussian curvature as a function of $\theta$ with $\phi=90^\circ$. Black and red points correspond to a magnetic field orientation in (c). Parameters: $t_\perp=3\times 10^{-3} eV$, $\eta=0$, $d/a=40$.}
\label{fig:alternative_geometry}
\end{figure*}

In the previous section, we considered the most natural geometry, where the crystallographic $\hat{c}$-axis is oriented along the $z$-axis. In this case, when $\theta=0$, the circular motion of an electron occurs in the $xy$ plane. However, cuprates can also be cut in a different way, with the copper planes oriented in the $yz$-plane, see Fig.~\ref{fig:alternative_geometry} (b). In this case, only the boundary points contribute to the Sondheimer oscillations. Fig.~\ref{fig:alternative_geometry} (c) shows a particular example of two such boundary points in the Brillouin zone. We refer to Appendix \ref{app:formalism} for details, and show the final formula for the conductivity below. The contribution to the conductivity from such points is given by

\begin{equation}
    \sigma_{\alpha \beta}^{\rm osc}=\frac{A_{\alpha \beta}(v_n^0)^2  (m^*)^2 \bar{v}_z^4}{2\pi^3 \omega_c^4 d^3 \left[\partial_{k_n} (m^* \bar{v}_z)\right]^2} e^{-\frac{d}{\bar{v}_z \tau}}\cos \left(\frac{ d \sqrt{K}}{\cos \theta}+\lambda_{\alpha \beta}\right)
    \label{eq:sondh_bound}
\end{equation}
where $K$ is the Gaussian curvature of the end point and the amplitude of conductivity decays as $B^{-4}$ with the magnetic field, as opposed to the $B^{-2.5}$ dependence discussed earlier. By rotating the magnetic field in the $yz$ plane and measuring the Sondheimer frequency, one can extract the Gaussian curvature of the whole Fermi surface, see Fig.~\ref{fig:alternative_geometry} (d).
We note, however, that the effective mass of these points is much larger than the effective mass in the previous geometry, since $t_z \ll (m_{x,y} a^2)^{-1}$. The oscillations require $\omega_c \tau \gg1$, and since the effective mass $m^*$ is enhanced, one needs much stronger magnetic fields to observe the Sondheimer oscillations.

Another interesting resonance occurs when the magnetic field is oriented along the crystallographic $\hat{c}$-axis, with $\theta=90^\circ$ and $\phi = 0$. In this case, the cyclotron motion of electrons happens in the copper $yz$ planes. If $d>r_c$, where $r_c$ is the cyclotron radius, and the disorder is weak, the electrons are bound within the planes and never reach the opposite ends of the sample. However, the resonance occurs when $d=r_c n$, and multiple electron orbits can fit in between two boundaries. The interference between the trajectories leads to an enhanced response on the opposite side of the film. Such resonances were studied by Kaner and Gantmakher \cite{Kaner_1968} in the context of anomalous penetration of the electromagnetic field inside the metal. The detailed analysis of this effect is beyond the scope of the current work, but it can be analyzed in the future.

\section{Discussion and conclusions}

In this section, we summarize our main results and discuss their relevance to real materials. We have proposed Sondheimer oscillations (SO) of the in-plane magneto-conductivity in thin films as a semiclassical transport probe of Fermi surface geometry in the cuprate pseudogap regime. Unlike quantum oscillations, SO arise from the commensuration between the cyclotron motion and the film thickness and therefore do not rely on Landau-level quantization and persist at relatively high temperatures where conventional (Shubnikov–de Haas) quantum oscillations are strongly suppressed. This makes SO a promising tool for accessing physics in a high-temperature pseudogap metal. Moreover, SO are sensitive to momentum-space trajectories for which $\partial^2 S/\partial k_z^2=0$, corresponding to regions of the Fermi surface with maximal out-of-plane velocity $\bar{v}_z$. In this sense, they provide complementary information to Shubnikov–de Haas oscillations, which instead probe extremal orbits satisfying $\partial S/\partial k_z=0$. This distinction can be particularly valuable in situations where the range of accessible magnetic field orientations is limited.

Our main result is a detailed calculation of the oscillatory conductivity for general quasi-2D Fermi surfaces within the Boltzmann framework. In the strong-field regime, the SO signal is controlled by two types of semiclassical trajectories in momentum space (corresponding to stationary phase contributions to Eq.~\eqref{eq:chambers_mod}): (i) non-contractible trajectories associated with extrema of the accumulated phase between the boundaries, which generically yield non-sinusoidal oscillations with a characteristic
power-law field dependence $\sim 1/B^{5/2}$, and (ii) boundary-point contributions that become important when the crystallographic $\hat{c}$-axis is perpendicular to the film orientation axis $\hat{n}$ and exhibit a $1/B^4$ field scaling. In both cases, the oscillation frequency is set by FS geometry (through $m^*$ and the averaged velocity $\bar{v}_z$ along the film axis $\hat{n}$ in (i), and through the Gaussian curvature $K$ in (ii)) and is independent of microscopic
scattering details. The amplitude is controlled by the ratio between the film's  thickness and the mean free path via an overall exponential factor.

Focusing on theoretical models relevant for cuprates, we computed the SO frequency spectrum
as a function of field orientation for three representative FS scenarios: an unreconstructed large FS, a spin-density-wave (SDW) reconstructed FS with pocket area $p/4$, and a fractionalized Fermi
liquid (FL$^*$) with pocket area $p/8$ (where $p$ is the hole doping). We identified several key features:
\begin{enumerate}
    \item Measuring the SO frequency $\Omega_{\rm SH}$ over field orientations provides direct access to FS structure. In particular, in the standard film geometry with $\hat{c}\parallel \hat{n}$, the in-plane
caliper radius of the FS can be extracted from a second derivative of $\Omega_{\rm SH}$ with respect to the angle $\theta$, see Eq.~(\ref{eq:caliper}). 
\item We also find that $\Omega_{\rm SH}$ exhibits pronounced features near Yamaji angles: the effective $\bar{v}_z$ controlling inter-layer motion is suppressed and the SO spectrum
shows sharp changes (see Fig.~\ref{fig:frequency_ancilla}), together with an exponential decay of oscillation amplitude. This naturally connects SO measurements to recent evidence for the Yamaji effect observed in angle-dependent magnetotransport experiments on Hg1201 \cite{Chan2024}. 

\item In contrast to non-oscillatory bulk magnetoresistance, distinct FS pockets contribute at distinct SH frequencies, so multiple pockets can, in principle, be separated in the Fourier spectrum. In addition, we find that the relative phase shift between oscillations in longitudinal and Hall conductivities depends on FS geometry as well: it takes a value $\pi/2$ for idealized elliptical pockets but can decrease substantially for a non-elliptical FS (such as arising in a FL$^*$ phase, see Fig.~\ref{fig:Fig_ancilla}(c)), allowing to further distinguish between candidate pseudogap scenarios.
\item For films cut such that  $\hat{c}\perp\hat{n}$, the dominant SO contribution is controlled by boundary points on the FS. In this geometry, the measured Sondheimer frequency is directly tied to the Gaussian curvature at those points, see Eq.~\eqref{eq:sondh_bound} and Fig.~\ref{fig:alternative_geometry}.
\item The evolution of $\Omega_{\rm SH}$ with doping across the overdoped–underdoped regime features a jump at the critical doping, see Fig.~\ref{fig:Fig_overdoped}. Such a jump,
if observed, would provide a  signature of a Fermi-volume-changing transition.
\end{enumerate}

Next, we discuss the magnitude of Sondheimer oscillations and whether it is possible to observe them in real materials. Sondheimer oscillations require two inequalities to hold \cite{gurevich1959}: $d < \bar{l}_z$ and $\omega_c \tau \gg1$. 
The first inequality arises from the exponential prefactor $e^{-d/\bar{l}_z}$ in Eqs. (\ref{eq:oscil_full_sp}), (\ref{eq:sondh_bound}). For parameters chosen in the caption of Fig. \ref{fig:Fig_1},\ref{fig:frequency_ancilla}, ten layers are enough to observe the amplitude of the order of a tenth of a percent. The second inequality is required to observe at least several oscillations. It particularly restricts the number of available materials, since in most cuprates strong fields are required to reach  $\omega_c \tau \gtrsim1$. The materials in which the regime $\omega_c \tau \gtrsim1$ has been accessed before are YBCO, Hg1201 and Tl2201.


Throughout the work, we use $c/a=4$, $\tau=0.55 ps$ and $d=40a$. For the band dispersion parameters, we  use the same parameters as in Ref.~\cite{Mascot2022}, see Appendix \ref{app:ancilla}. The elliptic approximation of the bands gives $m^*=1.66 m_e$ which is close to the values determined from quantum oscillations in the overdoped side of Tl2201\cite{LIU1992,Vignolle2008}. The Fermi energy of the hole pockets is $50 \,meV$, consistent with STM data. Finally, we assume that the hopping in the $z$ direction is the smallest parameter and choose $t_\perp=3\,meV$ and $t_\perp=6\,meV$ through this work. The calculations of the amplitude for an elliptical and FL$^*$ models are shown in Fig.~\ref{fig:Fig_1}(c) and Fig.~\ref{fig:Fig_ancilla}(c). In both cases, the amplitude is on the order of 0.5 percent compared to bulk conductivity. This makes such a measurement challenging, but doable.

In summary, our results demonstrate that Sondheimer oscillations provide a versatile and largely unexplored probe of Fermi-surface reconstruction in the cuprates. Systematic measurements of their angular dependence in underdoped and overdoped films could help discriminate between competing theoretical proposals and quantify Fermi-surface volumes. We also expect that Sondheimer oscillations could be studied
in other layered correlated materials where quantum oscillation measurements are challenging.

\begin{acknowledgements}
We are grateful to Brad Ramshaw, Bertrand Halperin, and Eslam Khalaf for useful discussions. This research was supported by NSF Grant DMR-2245246 and by the Simons Collaboration on Ultra-Quantum Matter which is a grant from the Simons Foundation (651440, S. S.). PAN is supported in part by a Harvard Quantum Initiative postdoctoral fellowship at Harvard University.
\end{acknowledgements}

\begin{widetext}

\appendix

\section{Derivation of Sondheimer oscillations for arbitrary Fermi surface}
\label{app:formalism}
In this appendix, we derive the Sondheimer conductivity given by Eq.~(\ref{eq:oscil_full_sp}) using the semiclassical Boltzmann approach. As explained in the main text, in the presence of a magnetic field, the linearized Boltzmann equation in the relaxation time approximation reads~\cite{Lifshitz1959}:
\begin{equation}
\omega_c\frac{\partial f_1}{\partial \varphi}+v_z \frac{\partial f_1}{\partial z} +\vec{E}\cdot\vec{v}=-\frac{f_1}{\tau}.
 \label{eqn:app_boltzman_z}
\end{equation}
 The fully diffusive boundary conditions translate into $f_1(z=0,v_z>0)=f_1(z=d,v_z<0)=0$. The equation can be simplified by doing a Fourier transform in the $z$-direction
$f_1(z,\varphi)=\sum_{n=-\infty}^{\infty} f_1(n,\varphi)e^{2 \pi i n z/d}$. Eq.~(\ref{eqn:app_boltzman_z}) becomes a first-order linear differential equation:

\begin{equation}
\omega_c\frac{\partial f_1(n ,\varphi)}{\partial \varphi}+f_1(n ,\varphi)\left(\frac{1}{\tau} +v_z \frac{2 \pi i n}{d} \right) = \vec{E}_n \cdot\vec{v}-v_z (f_1(d,\varphi)-f_1(0,\varphi)),
\end{equation}
where in general case $\vec{E}(z)=\sum_n \vec{E}_ne^{2 \pi i n z/d}$.
The solution is

\begin{equation}
    f_1(n,\varphi)=\frac{1}{\omega_c} \int_{-\infty}^\varphi d\varphi' E_{n \beta}v_\beta \exp \left(\int_\varphi^{\varphi'}d \varphi''(\gamma + i \kappa n)\right)-\frac{2}{\omega_c d} \int_{-\infty}^\varphi d\varphi' \sum_{n'} f_1(n' ,\varphi')|v_z| \exp \left(\int_\varphi^{\varphi'}d \varphi''(\gamma + i \kappa n)\right) ,  
    \label{eq:app_distr_funct}
    \end{equation}
where $\gamma =1/(\omega_c \tau)$ and $\kappa=2 \pi v_z /(d \omega_c)$. In full generality $\tau$ depends on $\varphi$ and $\gamma(\varphi)$ is inside the integral. The equation above is an integro-differential equation, which is nontrivial to solve. We first introduce the function $g(\varphi)= 2 \omega_c \sum_n f_1(n,\varphi)$ and solve for $g(\varphi)$. After summation, Eq.~(\ref{eq:app_distr_funct}) becomes

\begin{equation}
    g(\varphi)=2 \sum_n \int_{-\infty}^\varphi d\varphi' E_{n \beta}v_\beta \exp \left(\int_\varphi^{\varphi'}d \varphi''(\gamma + i \kappa n)\right)-\frac{2}{\omega_c d} \sum_n \int_{-\infty}^\varphi d\varphi' g(\varphi')|v_z| \exp \left(\int_\varphi^{\varphi'}d \varphi''(\gamma + i \kappa n)\right)    
    \end{equation}

We simply denote the first term as $2\Phi(\varphi)$. It is possible to show, see Ref.~\cite{gurevich1959}, that the solution of this integral equation depends on whether $v_z(\varphi)$ changes sign along the trajectory. If the sign of $v_z(\varphi)$ remains unchanged, the solution is
\begin{equation}
    g(\varphi)=\Phi(\varphi)-\Phi(\varphi_1)\exp(-\int_{\varphi_1(\varphi)}^\varphi \gamma d \varphi'), \quad \int_{\varphi_1(\varphi)}^\varphi v_z(\varphi')d\varphi'=\pm d \omega_c=\pm u \bar{v}_z,
\end{equation}
where $\bar{v}_z=(1/2 \pi)\int_0^{2\pi}v_z(\varphi)d\varphi$ is the average velocity in the $z$-direction. The variable $u=d \omega_c/\bar{v}_z$ will be useful later. Physically, $\varphi-\varphi_1(\varphi)$ corresponds to a phase winding when an electron travels between the two planes. In the latter case, when $v_z(\varphi)$ changes sign and the solution returns to the same plane before reaching the other one, the solution reads:
\begin{equation}
    g(\varphi)=\Phi(\varphi)-\Phi(\varphi_0)\exp(-\int_{\varphi_0(\varphi)}^\varphi \gamma d \varphi'), \quad \int_{\varphi_0(\varphi)}^\varphi v_z(\varphi')d\varphi'=0.
\end{equation}
We note that this type of solution does not contribute to Sondheimer oscillations and will be omitted later. 
The average current is given by:
\begin{equation}
    j_\alpha=\frac{1}{d}\int_0^d
j_\alpha(z) dz=\int  \frac{m^*}{8 \pi^3}v_\alpha(\varphi) f_1(0,\varphi) \frac{\partial f_0}{\partial \epsilon }  dk_n d\epsilon d \varphi 
\end{equation}
Using Eq.~(\ref{eq:app_distr_funct}) we obtain:

\begin{equation}
    j_\alpha=\int \frac{m^*}{(2\pi)^3}v_\alpha(\varphi)  \frac{\partial f_0}{\partial \epsilon }  dk_n d\epsilon d \varphi \left(\int_{-\infty}^\varphi \frac{1}{\omega_c}  d\varphi' E_{0 \beta}v_\beta(\varphi') \exp \left(\int_\varphi^{\varphi'}\gamma d \varphi'' \right) -\frac{1}{\omega_c^{2}d} \int_{-\infty}^\varphi d\varphi' g(\varphi')|v_z(\varphi')| \exp \left(\int_\varphi^{\varphi'}\gamma d \varphi'' \right)   \right)
    \label{eq:app_current}
\end{equation}
Eq.~(\ref{eq:app_current}) provides the most complete expression for the current. From this moment, let us assume that the electric field is constant, meaning only the $E_{0 \beta}$ component is present and $\tau(\varphi)=\tau$ does not depend on $\varphi$. Let us interpret three main contributions to Eq.~(\ref{eq:app_current}).

The first term is a three-dimensional Chambers formula \cite{Chambers_1952}:
\begin{equation}
    \sigma_{\alpha \beta}^{3D}=\int  \frac{m^*}{(2\pi)^3\omega_c} v_\alpha(\varphi)  \frac{\partial f_0}{\partial \epsilon }  dk_n d\epsilon d \varphi  \int_{-\infty}^\varphi d\varphi' v_\beta(\varphi') \exp \left(\gamma(\varphi'-\varphi)  \right)
    \label{cond:3d}
\end{equation}
In the limit of zero magnetic field ($B=0$) the Chambers formula can be elegantly rewritten as:
\begin{equation}
    \sigma_{\alpha \beta}^{3D}(B=0)= \int  \frac{m^* \tau}{(2\pi)^3} \frac{\partial f_0}{\partial \epsilon }  dk_n d\epsilon   \int_{0}^{2\pi}  v_\alpha(\varphi) v_\beta(\varphi)d \varphi 
    \label{cond:3d_B=0}
\end{equation}
The second term is a non-oscillatory correction, coming from the finite thickness and leading to an increased resistance in thin films \cite{Fuchs1938}. 
\begin{equation}
    \sigma_{\alpha \beta}^{corr}=- \int \frac{m^*}{(2\pi)^3 \omega_c^{2}d } v_\alpha(\varphi)  \frac{\partial f_0}{\partial \epsilon }  dk_n d\epsilon d \varphi  \int_{-\infty}^\varphi d\varphi' |v_z(\varphi')|  \int_{-\infty}^{\varphi'}d\varphi'' v_\beta(\varphi'') \exp \left(\gamma(\varphi''-\varphi)  \right)
\end{equation}
The nontrivial oscillatory contribution to the conductivity comes from the last term:
\begin{equation}
    \sigma_{\alpha \beta}^{\rm osc}=  \int \frac{m^*}{(2\pi)^3 \omega_c^{2}d }  v_\alpha(\varphi)  \frac{\partial f_0}{\partial \epsilon }  dk_n d\epsilon d \varphi  \int_{-\infty}^\varphi d\varphi'  |v_z(\varphi')|\int_{-\infty}^{\varphi_1(\varphi')} d\varphi'' v_\beta(\varphi'')   \exp \left(\gamma(\varphi''-\varphi)  \right)
    \label{eqn:app_cond_gen}
\end{equation}
It was first derived by Sondheimer \cite{Sondheimer1952} for a spherical Fermi surface. He showed that the oscillations are linear in the magnetic field $B$ and decay as $B^{-4}$ with increasing the magnetic field. However, the $B^{-4}$ dependence is non-universal as was shown later by Gurevich \cite{gurevich1959}, and depends on the shape of the Fermi surface, see the discussion below.

From now on, we focus on the oscillating component of the conductivity, given by Eq.~(\ref{eqn:app_cond_gen}). This formula can be written compactly as:
\begin{equation}
    \sigma_{\alpha \beta}^{\rm osc}=\int  \frac{m^*}{(2\pi)^3 \omega_c^{2}d } d k_n d \epsilon \frac{\partial f_0}{\partial \epsilon }   B_{ \alpha \beta}(k_n,u),
    \label{eq:oscil_full}
\end{equation}
where $u=d \omega_c/\bar{v}_z$ was defined earlier.
To make further analytical progress, we differentiate $B_{ \alpha \beta}(k_n,u)$ twice with respect to $u$. Using $\partial \varphi_1(\varphi,u)/\partial u=-\bar{v}_z/v_z(\varphi_1(\varphi))$:
\begin{equation}
    \frac{\partial^2 B_{\alpha \beta}(k_n,u)}{\partial u^2}=   \int \bar{v}_z^2   d \varphi    \frac{  v_\alpha(\varphi) v_\beta(\varphi_1(\varphi))  }{v_z(\varphi_1(\varphi))} \exp \left(\gamma(\varphi_1(\varphi)-\varphi)  \right)\;.
    \label{eq:oscil_full_2}
\end{equation}
We can further approximate $\varphi_1(\varphi)=\varphi-\omega_c d/\bar{v}_z$ and  $\gamma(\varphi_1(\varphi)-\varphi) \approx -d/(\bar{v}_z \tau)= -u /(\omega_c \tau)$.
Then, the resulting expression for the second derivative reads:
\begin{equation}
  \frac{\partial^2 B_{\alpha \beta}(k_n,u)}{\partial u^2}=e^{-u /(\omega_c \tau)}   \int \bar{v}_z^2   d \varphi    \frac{  v_\alpha(\varphi) v_\beta(\varphi_1(\varphi))  }{v_z(\varphi_1(\varphi))} =e^{-u /(\omega_c \tau)}  c_{ \alpha \beta}(k_n,u)\;.
    \label{eq:oscil_full_3}
\end{equation}
We note that $c_{\alpha \beta }(k_n,u)$ is a periodic functions of $u$ with period $2\pi$ which follows from the fact that $v_\alpha(\varphi)=v_\alpha(\varphi+2 \pi n)$. Therefore, without loss of generality, $B_{\alpha \beta}(k_n,u)=e^{-u /(\omega_c \tau)} b_{ \alpha \beta}(k_n,u)$ where $b_{ \alpha \beta}(k_n,u)$ is a periodic function of $u$ with period $2\pi$. At zero temperature, Eq.~(\ref{eq:oscil_full}) reads
 \begin{equation}
 \sigma_{\alpha \beta}^{\rm osc}=  \int  \frac{m^*}{(2 \pi)^3 \omega_c^{2}d}  dk_n  e^{-u/(\omega_c \tau)}b_{\alpha \beta}(k_n,u) \;.
    \label{eq:app_oscil_full_4}
\end{equation}
Since $b_{\alpha \beta}(k_n,u)$ is a periodic function of $u$, the integral oscillates rapidly at strong magnetic fields. Hence, the stationary point method can be applied to evaluate the integral. The first contribution comes from the boundary points where $k_n$ reaches extremal value and trajectories converge to a point in the Brillouin zone. They lead to $B^{-4}$ amplitude decay, as we will show in the next section. The second contribution is coming from stationary points $k^*$ where the integral stops oscillating $du(k^*)/dk_n=0$.

Since $b_{\beta\alpha}(k_n,u)$ is periodic we can always perform a Fourier expansion:
\begin{equation}
    b_{\alpha \beta}(k_n,u)=\sum_s e^{i s u}b^{(s)}_{\alpha \beta}(k_n), \quad u(k_n)=u^*+(1/2)u''(k^*)(k_n-k^*)^2\;.
\end{equation}
The Fourier harmonics $b^{(s)}_{\alpha \beta}(k_n)$ and $c^{(s)}_{\alpha \beta}(k_n)$ are related via, see Eq.~(\ref{eq:oscil_full_3}):
\begin{equation}
    b^{(s)}_{\alpha \beta}(k_n)=\frac{c^{(s)}_{\alpha \beta}(k_n)}{(i s-1/(\omega_c\tau))^2}\;.
\end{equation}
Using the stationary point approximation and integrating over $k_n$ near the stationary point $k^*$ we obtain:

 \begin{equation}
 \sigma_{\alpha \beta}^{\rm osc}=  \frac{m^* \tau^2}{(2\pi)^3d} e^{-d/(\bar{v}_z \tau)}\sum_{s=-\infty}^{+\infty} \frac{ c^{(s)}_{\alpha \beta}(k^*)}{(1- i s \, \omega_c \tau)^2}  \sqrt{\frac{2 \pi i }{s \,u''(k^*)}}e^{i s \Omega_{\rm SH} B}\Bigg|_{k_n=k^*}\;,  
    \label{eq:app_oscil_full_sp}
\end{equation}
where $\Omega_{\rm SH}=d/(m^* \bar{v}_z)$ is Sondheimer frequency. We note that in general case the oscillations are not described by a single harmonic, and could be non-sinusoidal. The amplitude of the oscillations is suppressed exponentially by a factor $e^{-d/(\bar{v}_z \tau)}$ and disappears in thick samples. The term
 $u'' \propto \omega_c$, therefore the amplitude of the oscillations decays as $B^{-2.5}$ as magnetic field increases. This distinguishes them from the other type of oscillations, coming from the boundary points, to which we turn our attention now.

Finally, let us briefly comment on the temperature effects. For simplicity, we restrict ourselves to a perpendicular magnetic field $\theta=0$ and an elliptical Fermi surface, given by Eq.~(\ref{eq:disp_elliptic}). Like any semiclassical effect described by Boltzmann theory, Sondheimer oscillations do not require temperature to be smaller than any energy scale associated with level quantization. Instead, the main condition we have implicitly assumed so far is $T\ll \epsilon_F$. Given that the Fermi pockets in the pseudogap phase are relatively small, and the temperature may be relatively high, it is natural to ask how sensitive our results are to a slight relaxation of this condition. Still assuming $ t_{\perp}\ll \epsilon_F$ but allowing for $T\sim \epsilon_F$, Eq.~(\ref{eq:oscil_full}) reads:
\begin{equation}
    \sigma_{xx}^{\rm osc}=\int  \frac{m^*}{(2\pi)^3 \omega_c^{2}d } d k_z d \epsilon \left(-\frac{\partial f_0}{\partial \epsilon }\right)  \frac{2 \pi \epsilon v_z }{m_x} \Re \left( e^{-u/\omega_c \tau+i u} \frac{\omega_c^2 \tau ^2}{(1-i \omega_c \tau)^2} \right)\;.
\end{equation}
The remaining frequency integral is decoupled and can be easily evaluated
\begin{equation}
I=\int  d \epsilon \frac{\partial f_0}{\partial \epsilon } \epsilon=\epsilon_F x \ln(1+e^{1/x}) ,\quad x=T/\epsilon_F\;.
\end{equation}
For $x=0.5$, approximately corresponding to room temperature, $I=1.06$, so the temperature corrections are insignificant. Although this is not directly relevant to the regime discussed here, we note that if the condition $T\ll t_\perp$ is violated as well, then the temperature effects become more subtle, and we refer to \cite{nikolaenko2025} for details.

\subsection{Contribution from the boundary points}

The second type of contribution to the oscillating conductivity in Eq.~(\ref{eq:app_oscil_full_4}) is coming from the boundary points $k^*$. In the following, we assume the $k^*$ is an elliptic point, and the trajectory of the electron converges to one point. The Fermi surface around the boundary point can be described by the following quadratic form in the basis $(x',y',n)$ where $n$ component is aligned with the magnetic field through the angles $(\theta, \phi)$: 
\begin{equation}
    v_n \Delta k_n+\frac{1}{2}\beta_{i j} \Delta k_i \Delta k_j=0\;, 
\end{equation}
where  $\beta_{ij}=\partial^2 \epsilon/(\partial k_i \partial k_j)$ is positive definite and $\Delta k_i,\Delta k_j \in (x',y')$ are the deviations from the $k^*$ in the plane perpendicular to magnetic field. The principal curvatures of the surface are  $\beta_1/v_n, \, \beta_2/v_n$, where $\beta_1,\beta_2$ are eigenvalues of $\beta_{ij}$   and the Gaussian curvature $K=\beta_1 \beta_2/v_n^2$.   The effective mass of the trajectory around the boundary point is given by $m^*=1/\sqrt{\beta_1 \beta_2}$. 

The equation of motion 
 $\omega_c  d \vec{k}/d\varphi= \vec{v} \times \vec{B}$ can be solved around $k^*$ and the resulting velocities are:
 \begin{equation}
     \begin{split}
         &v_x''(\varphi)=\sqrt{2 v_n^0 \Delta k_n\beta_1} \cos \varphi\\
                  &v_y''(\varphi)=\sqrt{2 v_n^0 \Delta k_n\beta_2} \sin \varphi\\
                           &v_n(\varphi)=v_n^0+\sqrt{2 v_n^0 \Delta k_n} \left(\beta_{zx'}\beta_1^{-1/2}\cos \varphi +\beta_{zy'}\beta_1^{-1/2}\sin \varphi\right),\\
     \end{split}
 \end{equation}
where $(x'',y'')$ are the principal axes of the quadratic form $\beta_{ij}$ and $n$ is the direction parallel to the magnetic field. After that, it is straightforward but cumbersome to calculate the velocities in the original basis. They can be obtained by applying two successive rotations: first from the principal basis $(x'',y'',n)$ to $(x',y',n)$, and subsequently by the angles $(\theta,\phi)$ to return to the original coordinate system. The resulting velocities are plugged into Eq.~(\ref{eq:oscil_full_3}), the integral is readily computed and is proportional to either $\cos u$ or $\sin u$. If the coefficients $\beta_{ij}$ are comparable, the resulting expression for oscillating part of $B_{\alpha \beta}$ acquires the form:
\begin{equation}
    B_{\alpha \beta}(u)=\frac{A_{\alpha \beta}}{m^*} e^{-u/(\omega_c \tau)} v_n^2 \Delta k_n \cos (u+\lambda_{\alpha \beta })\;,
\end{equation}
Where $A_{\alpha \beta}$ and $\lambda_{\alpha \beta}$ depend on the difference in coefficients $\beta_{ij}$ and on the angles $(\theta, \phi, \phi')$ of initial and principal value axes. In most cases  $A_{\alpha \beta}$ is of order of unity and is not important for further discussion.

Integrating Eq.~(\ref{eq:app_oscil_full_4}) by parts two times around boundary point we obtain:
\begin{equation}
    \sigma_{\alpha \beta}^{\rm osc}=\frac{A_{\alpha \beta}(v_n^0)^2e^{-d/(\bar{v}_z \tau)}}{2\pi^3 \omega_c^4 d^3 (m^*)^2 \left[d(1/(m^* \bar{v}_z))/dk_n\right]^2} \cos \left( \frac{d \omega_c }{\bar{v}_z}+\lambda_{\alpha \beta}\right)\;.
\end{equation}
In this case the Sondheimer frequency is given by $\Omega_{\rm SH}=d/(m^* v_n^0 \cos \theta)=d \sqrt{K}/\cos \theta$, where $K$ is the Gaussian curvature. The oscillation amplitude decays exponentially with the thickness of the film and as $B^{-4}$ with the magnetic field. The different power law decay distinguishes boundary point contributions from the stationary points.

\section{Application to the quasi-2D elliptic Fermi surface}

In this appendix, we focus on the elliptic Fermi surface, where most results can be derived analytically. The dispersion of the electron is 
\begin{equation}
    \epsilon=\frac{k_x^2}{2m_x}+\frac{k_y^2}{2m_y}-2 t_\perp \cos (c k_z)\;,
\end{equation}
and we assume that $t_z \ll (m_{x,y} a^2)^{-1}$, where $a$ is the lattice constant in the $xy$ plane.
The semiclassical equation of motion describing the electron in a magnetic field is written
\begin{equation}
    \frac{d \mathbf{k}}{dt}= \mathbf{v} \times \mathbf{B}, \quad \mathbf{v}=\partial_{\mathbf{k}} E(\mathbf{k})\;.
\end{equation}

Introducing the magnetic field $\mathbf{B}=B(\sin \theta \cos \phi, \sin \theta \sin \phi, \cos \theta)$ the equations read:

 \begin{equation}
      \begin{cases}
\dfrac{d k_x}{d t} = 
  \dfrac{B}{m_y} \cos\theta \, k_y 
  - 2 t_\perp  B \sin k_z \, \sin\theta \, \sin\phi, \\[1.2em]

\dfrac{d k_y}{d t} = 
  -\dfrac{B}{m_x} \cos\theta \, k_x 
  + 2 t_\perp  B \sin k_z \, \sin\theta \, \cos\phi, \\[1.2em]

\dfrac{d k_z}{d t} = 
  \dfrac{B}{m_x} k_x \, \sin\theta \, \sin\phi 
  - \dfrac{B}{m_y} k_y \, \sin\theta \, \cos\phi,
\end{cases}
\end{equation}
where $v_{x}=k_{x}/m_{x}$, $v_{y}=k_{y}/m_{y}$  and $v_z=2 t_\perp c\sin(c k_z)$. We solve these equations by neglecting $t_\perp$ to first order:
\begin{equation}
    k_x=\sqrt{2 m_x \epsilon_F} \cos\left(\frac{B \cos \theta}{\sqrt{m_x m_y}}t+\phi_0\right), \,  k_y=-\sqrt{2 m_y \epsilon_F} \sin\left(\frac{B \cos \theta}{\sqrt{m_x m_y}}t+\phi_0\right)\;.
\end{equation}
Thus, the effective mass $m^*=\sqrt{m_x m_y}/\cos \theta$ does not depend on the momenta and the cyclotron frequency is $\omega_c=B/m^*$.
The last equation for $k_z$ can be solved and the average $\bar{v}_z$ becomes

\begin{equation}
    \bar{v}_z= 2 t_\perp c \sin (c k_{z0})J_0(c\sqrt{2 \epsilon_F(m_x \cos^2 \phi+m_y \sin ^2 \phi) } \tan \theta)\;.
    \label{eq:sondh_analytic}
\end{equation}
The stationary points of $u=d/(m^* \bar{v}_z)$ occur at $ck_{z0}=\pi/2$ and the Sondheimer frequency is equal to 
\begin{equation}
 \Omega_{\rm SH}(\theta,\phi)=\frac{d  \cos \theta}{2 t_\perp c \sqrt{m_x m_y} J_0\left(c\sqrt{2 \epsilon_F(m_x \cos^2 \phi+m_y \sin ^2 \phi) } \tan \theta \right)}\;,
\end{equation}
compare to Eq.~(\ref{eq:sondh_ell}) of the main text. The variable inside the Bessel function $k_{\rm cal}(\phi)=\sqrt{2 \epsilon_F(m_x \cos^2 \phi+m_y \sin ^2 \phi) } $ is the caliper radius of an ellipse.

\section{Details of the dispersion relation in FL$^*$ and SDW models}
\label{app:ancilla}

In this Appendix, we provide more information about the  dispersion relations used to describe the pseudogap phase. The effective Hamiltonian for the FL$^*$ is given by Eq.~(\ref{eq:FLs}), where $\epsilon_c(\mathbf{k})$ is a dispersion of physical electrons, $\epsilon_f(\mathbf{k})$ is a dispersion of hidden electrons, and $\Phi$ is a hybridization between two Ancilla layers.
The dispersion for the electron layer is given by a tight-binding model:
\begin{equation}
\begin{aligned}
\epsilon_c(\mathbf{k}) =&-2t(\cos k_x+\cos k_y)-4 t' \cos k_x \cos k_y-2t''(\cos 2 k_x+\cos 2 k_y) \nonumber \\
&-4t'''(\cos 2 k_x \cos k_y+\cos 2 k_y \cos k_x)-\mu_c\,,
\end{aligned}
\end{equation}
where $t=0.22, t'=-0.034, t''=0.036, t'''=-0.007, \mu_c=-0.24$ (all energies are in units of eV). The parameters used here were used to fit the overdoped side of BI2201, see \cite{Mascot2022,He2011} for details.

The hidden layer dispersion is also given by a tight-binding model with dispersion
\begin{equation}
	\epsilon_f(\mathbf{k}) = 2t_1(\cos k_x+\cos k_y)
	+4 t_1' \cos k_x \cos k_y +2t_1''(\cos 2 k_x+\cos 2 k_y)-\mu_f\,, \label{eq:ef}
\end{equation}
where the parameters are $t_1=0.1, t_1'=-0.03, t_1''=-0.01, \mu_f=0.009$. Finally, we use the hybridization $\Phi=0.09$. In the presence of hopping in the $z$-direction, the dispersion reads:

\begin{equation}
\begin{split}
 & E_{\pm}(\mathbf{k})=\frac{\epsilon_c(\mathbf{k})+\epsilon_f(\mathbf{k})-2 t_\perp \cos (c k_z)}{2} \\
 &\pm \sqrt{\left(\frac{\epsilon_c(\mathbf{k})-\epsilon_f(\mathbf{k})}{2}\right)^2+\Phi^2 +t_\perp^2 \cos (c k_z)^2-t_\perp \cos (c k_z)(\epsilon_c(\mathbf{k})-\epsilon_f(\mathbf{k})) \cos(2\eta)-2 t_\perp \cos(c k_z) \Phi \sin (2\eta)}\;.\\
\label{eq:Epm_tz}  
\end{split}
\end{equation}





\end{widetext}

\bibliography{bibliography}

\end{document}